\documentclass[conference]{IEEEtran}

\usepackage[nolist]{acronym} 
\usepackage{algorithm,algpseudocode}
\usepackage{pgfplots}
\usepackage{tablefootnote}
\usepgfplotslibrary{groupplots}
\usepackage{graphicx}
\pgfplotsset{compat=newest}
\usepackage[]{nohyperref}  % This makes hyperref commands do nothing without errors
\usepackage{units}
\usepackage{amsmath, amsbsy, amssymb}
\usepackage{tikzscale}
\usepackage{color}
\usepackage{epigraph}
\usepackage{circuitikz}
\usepackage[group-separator={,}]{siunitx}
\usetikzlibrary{matrix}
\usetikzlibrary{positioning}
\usepackage{bbm}

\definecolor{mittelblau}{RGB}{0, 126, 198}
\definecolor{violettblau}{cmyk}{0.9, 0.6, 0, 0}
\definecolor{rot}{RGB}{238, 28 35}
\definecolor{apfelgruen}{RGB}{140, 198, 62}
\definecolor{gelb}{RGB}{1, 221, 0}
\definecolor{orange}{RGB}{244, 111, 33}
\definecolor{pink}{RGB}{237, 0, 140}
\definecolor{lila}{RGB}{128, 10, 145}
\definecolor{hellgrau}{RGB}{224, 224, 224}
\definecolor{mittelgrau}{RGB}{128, 128, 128}
\definecolor{dunkelgrau}{RGB}{80,80,80}
\definecolor{anthrazit}{RGB}{19, 31, 31}

\definecolor{myblue}{RGB}{80,80,160} 
\definecolor{mygreen}{RGB}{80,160,80}
\definecolor{myorgange}{RGB}{204,102,0}
\definecolor{lightblue}{RGB}{51,153,255}

\usetikzlibrary{shapes,arrows,spy}
\usetikzlibrary{dsp}

\IEEEoverridecommandlockouts

\newcommand\deact[1]{}

\begin{document}

%\title{Exploiting Universality of Spatially Coupled LDPC Codes for Multi-User IDMA Systems}
\title{Spatially Coupled LDPC Codes and the \\ Multiple Access Channel}

\author{\textit{(Invited Paper)}\\\quad\\
\IEEEauthorblockN{Sebastian Cammerer, Xiaojie Wang, Yingyan Ma, and Stephan ten Brink}

\IEEEauthorblockA{
 Institute of Telecommunications, Pfaffenwaldring 47, University of  Stuttgart, 70569 Stuttgart, Germany 
   \\\{cammerer,wang,tenbrink\}@inue.uni-stuttgart.de
%\\
%\IEEEauthorrefmark{2}Nokia Bell Labs, Lorenzstr. 10, 70435 Stuttgart, Germany, %laurent.schmalen@nokia-bell-labs.com
}

}

\maketitle

% Vectors
\renewcommand{\vec}[1]{\mathbf{#1}}
\newcommand{\vecs}[1]{\boldsymbol{#1}}

\newcommand{\av}{\vec{a}}
\newcommand{\bv}{\vec{b}}
\newcommand{\cv}{\vec{c}}
\newcommand{\dv}{\vec{d}}
\newcommand{\ev}{\vec{e}}
\newcommand{\fv}{\vec{f}}
\newcommand{\gv}{\vec{g}}
\newcommand{\hv}{\vec{h}}
\newcommand{\iv}{\vec{i}}
\newcommand{\jv}{\vec{j}}
\newcommand{\kv}{\vec{k}}
\newcommand{\lv}{\vec{l}}
\newcommand{\mv}{\vec{m}}
\newcommand{\nv}{\vec{n}}
\newcommand{\ov}{\vec{o}}
\newcommand{\pv}{\vec{p}}
\newcommand{\qv}{\vec{q}}
\newcommand{\rv}{\vec{r}}
\newcommand{\sv}{\vec{s}}
\newcommand{\tv}{\vec{t}}
\newcommand{\uv}{\vec{u}}
\newcommand{\vv}{\vec{v}}
\newcommand{\wv}{\vec{w}}
\newcommand{\xv}{\vec{x}}
\newcommand{\yv}{\vec{y}}
\newcommand{\zv}{\vec{z}}
\newcommand{\zerov}{\vec{0}}
\newcommand{\onev}{\vec{1}}
\newcommand{\alphav}{\vecs{\alpha}}
\newcommand{\betav}{\vecs{\beta}}
\newcommand{\gammav}{\vecs{\gamma}}
\newcommand{\lambdav}{\vecs{\lambda}}
\newcommand{\omegav}{\vecs{\omega}}
\newcommand{\sigmav}{\vecs{\sigma}}
\newcommand{\tauv}{\vecs{\tau}}

% Matrices
\newcommand{\Am}{\vec{A}}
\newcommand{\Bm}{\vec{B}}
\newcommand{\Cm}{\vec{C}}
\newcommand{\Dm}{\vec{D}}
\newcommand{\Em}{\vec{E}}
\newcommand{\Fm}{\vec{F}}
\newcommand{\Gm}{\vec{G}}
\newcommand{\Hm}{\vec{H}}
\newcommand{\Id}{\vec{I}}
\newcommand{\Jm}{\vec{J}}
\newcommand{\Km}{\vec{K}}
\newcommand{\Lm}{\vec{L}}
\newcommand{\Mm}{\vec{M}}
\newcommand{\Nm}{\vec{N}}
\newcommand{\Om}{\vec{O}}
\newcommand{\Pm}{\vec{P}}
\newcommand{\Qm}{\vec{Q}}
\newcommand{\Rm}{\vec{R}}
\newcommand{\Sm}{\vec{S}}
\newcommand{\Tm}{\vec{T}}
\newcommand{\Um}{\vec{U}}
\newcommand{\Vm}{\vec{V}}
\newcommand{\Wm}{\vec{W}}
\newcommand{\Xm}{\vec{X}}
\newcommand{\Ym}{\vec{Y}}
\newcommand{\Zm}{\vec{Z}}
\newcommand{\Lambdam}{\vecs{\Lambda}}
\newcommand{\Pim}{\vecs{\Pi}}

% Calligraphic
\newcommand{\Ac}{{\cal A}}
\newcommand{\Bc}{{\cal B}}
\newcommand{\Cc}{{\cal C}}
\newcommand{\Dc}{{\cal D}}
\newcommand{\Ec}{{\cal E}}
\newcommand{\Fc}{{\cal F}}
\newcommand{\Gc}{{\cal G}}
\newcommand{\Hc}{{\cal H}}
\newcommand{\Ic}{{\cal I}}
\newcommand{\Jc}{{\cal J}}
\newcommand{\Kc}{{\cal K}}
\newcommand{\Lc}{{\cal L}}
\newcommand{\Mc}{{\cal M}}
\newcommand{\Nc}{{\cal N}}
\newcommand{\Oc}{{\cal O}}
\newcommand{\Pc}{{\cal P}}
\newcommand{\Qc}{{\cal Q}}
\newcommand{\Rc}{{\cal R}}
\newcommand{\Sc}{{\cal S}}
\newcommand{\Tc}{{\cal T}}
\newcommand{\Uc}{{\cal U}}
\newcommand{\Wc}{{\cal W}}
\newcommand{\Vc}{{\cal V}}
\newcommand{\Xc}{{\cal X}}
\newcommand{\Yc}{{\cal Y}}
\newcommand{\Zc}{{\cal Z}}

\newcommand{\CN}{\Cc\Nc}

% Number sets
\newcommand{\CC}{\mathbb{C}}
\newcommand{\MM}{\mathbb{M}}
\newcommand{\NN}{\mathbb{N}}
\newcommand{\RR}{\mathbb{R}}

% Mixed symbols
\newcommand{\htp}{^{\mathsf{H}}}
\newcommand{\tp}{^{\mathsf{T}}}

% Brackets
\newcommand{\LB}{\left(}
\newcommand{\RB}{\right)}
\newcommand{\LP}{\left\{}
\newcommand{\RP}{\right\}}
\newcommand{\LSB}{\left[}
\newcommand{\RSB}{\right]}

\renewcommand{\ln}[1]{\mathop{\mathrm{ln}}\LB #1\RB}
\newcommand\norm[1]{\left\lVert#1\right\rVert}
\newcommand{\cs}[1]{\mathop{\mathrm{cs}}\LSB #1\RSB}

% Expectation, Variance, etc
\newcommand{\EE}{{\mathbb{E}}}
\newcommand{\Expect}[2]{\EE_{#1}\LSB #2\RSB}

% Theorems, Lemma, etc.
\newtheorem{definition}{Definition}[section]
\newtheorem{remark}{Remark}

\begin{acronym}
 \acro{ADC}{analog-to-digital converter}
 \acro{AGC}{automatic gain control}
 \acro{ASIC}{application-specific integrated circuit}
 \acro{AWGN}{additive white Gaussian noise}
 \acro{BER}{bit error rate}
 \acro{BICM}{bit interleaved coded modulation}
 \acro{BLER}{block error rate}
 \acro{CFO}{carrier frequency offset}
 \acro{DL}{deep learning}
 \acro{DQPSK}{differential quadrature phase-shift keying}
 \acro{ECC}{error correcting code}
 \acro{FPGA}{field programmable gate array}
 \acro{GNR}{GNU Radio}
 \acro{GPU}{graphic processing unit}
 \acro{ISI}{inter-symbol interference}
 \acro{LOS}{line-of-sight}
 \acro{MIMO}{multiple-input multiple-output}
 \acro{ML}{machine learning}
 \acro{MLP}{multilayer perceptron}
 \acro{MSE}{mean squared error}
 \acro{NN}{neural network}
 \acro{PLL}{phase-locked loop}
 \acro{ppm}{parts per million}
 \acro{PSK}{phase-shif keying}
 \acro{PFB}{polyphase filterbank}
 \acro{QAM}{quadrature amplitude modulation}
 \acro{ReLU}{rectified linear unit}
 \acro{RNN}{recurrent neural network}
 \acro{RRC}{root-raised cosine}
 \acro{RTN}{radio transformer network}
 \acro{SDR}{software-defined radio}
 \acro{SFO}{sampling frequency offset}
 \acro{SGD}{stochastic gradient descent}
 \acro{SNR}{signal-to-noise ratio}	
 \acro{TDL}{tapped delay line}
 \acro{OFDM}{orthogonal frequency division multiplex}
 \acro{IFFT}{inverse fast Fourier transform}
 \acro{FFT}{fast Fourier transform}
 \acro{IFT}{inverse Fourier transform}
 \acro{FT}{Fourier transform}
 \acro{IDFT}{inverse discrete Fourier-transform}
 \acro{DFT}{discrete Fourier-transform}
 \acro{CP}{cyclic prefix}
 \acro{MMSE}{minimum mean squared error}
 \acro{QPSK}{quadrature phase-shift keying}
 \acro{BP}{belief propagation}
 \acro{SC}{spatial coupling}
 \acro{LDPC}{low-density parity-check}
 \acro{SC-LDPC}{spatially coupled low-density parity-check}
 \acro{DE}{density evolution}
 \acro{MAP}{maximum a posteriori}
 \acro{VN}{variable node}
 \acro{CN}{check node}
 \acro{LLR}{log likelihood ratio}
 \acro{NOMA}{non-orthogonal multiple access}
 \acro{IDMA}{interleave division multiple access}
 \acro{MUD}{multi-user detector}
 \acro{VND}{variable node decoder}
 \acro{CND}{check node decoder}
 \acro{GA}{Gaussian approximation}
 \acro{EXIT}{extrinsic information transfer}
 \acro{BPSK}{binary phase shift keying}
 \acro{GMAC}{Gaussian multiple access channel}
 \acro{REP}{repetition code}
\end{acronym}

\begin{abstract}
We consider \ac{SC-LDPC}\acused{LDPC} codes within a non-orthogonal \ac{IDMA} scheme to avoid cumbersome degree profile \emph{matching} of the \ac{LDPC} code components to the iterative \ac{MUD}.
Besides excellent decoding thresholds, the approach benefits from the possibility of using rather simple and regular underlying block \ac{LDPC} codes owing to the \emph{universal} behavior of the resulting coupled code with respect to the channel front-end, i.e., the iterative \ac{MUD}.
Furthermore, an additional outer repetition code makes the scheme flexible to cope with a varying number of users and user rates, as the \ac{SC-LDPC} itself can be kept constant for a wide range of different user loads.
%To avoid any re-design of the \ac{SC-LDPC} code when the number of users changes, we show that the concatenation of an outer repetition code allows high flexibility by adjusting the repetition rate whenever the number of users in the system changes.
%We analyze the decoding thresholds via \ac{DE} and verify its correctness by corresponding \ac{BER} simulations.
The decoding thresholds are obtained via \ac{DE} and verified by \ac{BER} simulations.
%As a result, we achieve decoding thresholds within XY dB to the Shannon capacity for $N=32$ users.
To keep decoding complexity and latency small, we introduce a joint iterative windowed detector/decoder imposing carefully adjusted sub-block interleavers.
Finally, we show that the proposed coding scheme also works for Rayleigh channels using the same code with tolerable performance loss compared to the \ac{AWGN} channel.
\end{abstract}

\acresetall

\section{Introduction}

While in the single-user case the Shannon capacity has been almost achieved for practical coding schemes \cite{BERROU93,chung2001design}, the situation changes when considering the multi-user \ac{NOMA} scenario, i.e., when multiple transmitters and a single receiver share the same medium \cite{DTseBookFund}. 
Although several \ac{NOMA} approaches exist (see \cite{NOMASurvey,NOMA5GCommMag} and references therein), it is still an open and interesting research direction to find low-complexity coding (and detection) schemes that operate close to the multi-user capacity. Further, multi-user systems open up yet unsettled research opportunities such as the flexibility towards a dynamically varying number of users and, likely, having different power levels.

One attractive \ac{NOMA} scheme, featuring low-complexity, parallelizable computation and asynchronous transmission, is \ac{IDMA} \cite{LiPingIDMACL04,LipingIDMATWC06}. In this work, we focus on the \ac{IDMA} scheme, where a low-complexity parallel interference cancellation (PIC) receiver is used  and an effective separation between users is done by an individual interleaver. Relying on feedback from single-user channel decoders, the performance of \ac{IDMA} systems strongly depends on the performance of the underlying channel codes.

Thus, in classical \ac{LDPC}-based \ac{IDMA} systems, the \ac{LDPC} code needs to have \emph{matched} degree profiles \cite{TB_KR_ASH_LDPC04,wang2018idma} to the channel front-end, i.e., the \ac{MUD}. As this optimization depends on several parameters such as the channel type, the number of users and the individual \acp{SNR}, the drawback is that in practice either multiple \ac{LDPC} codes need to be (pre-)designed, or a degraded system performance has to be accepted.
In this work, we make use of the fact that \ac{SC-LDPC} codes do not have this drawback as they are known for a \emph{universal} behavior regarding the channel front-end \cite{Schmalentenbrink, cammerer2016wave}.

\ac{SC-LDPC} codes are widely known for their capacity achieving decoding behavior via \emph{threshold saturation} \cite{Coupl11BMS}. More precisely, it has been shown that coupled codes approach the \ac{MAP} decoding performance of the underlying block \ac{LDPC} code under low-complexity \ac{BP} decoding for properly chosen code parameters. 
However, a second powerful property of \ac{SC-LDPC} codes is not so often referred to, but can be seen in their \emph{universality} with respect to the channel front-end, i.e., for carefully chosen code parameters \ac{SC-LDPC} codes do not need any re-design when the channel characteristics change. It also offers potentially low error-floors as typically a regular code design suffices which simplifies the code construction.
This universality renders \ac{SC-LDPC} codes into a promising candidate for \ac{NOMA} schemes where a wide range of different scenarios must be supported such as different number of users, channels and user power.

Unfortunately, as \ac{SC-LDPC} codes are constructed out of multiple coupled sub-blocks, they typically introduce long block lengths. Thus, for practical decoder implementations, i.e., feasible decoding complexity, a windowed decoder is crucial \cite{iyengar2013windowed}. We show that the \ac{MUD} can be integrated into the iterative detection/decoding scheme with negligible performance loss. Yet, this requires some attention with respect to the decoding window and the interleaver design to separate individual users, as the iterative detection/decoding loop needs to be performed sub-block-wise to maintain the benefits of threshold saturation.

\section{IDMA System Model}

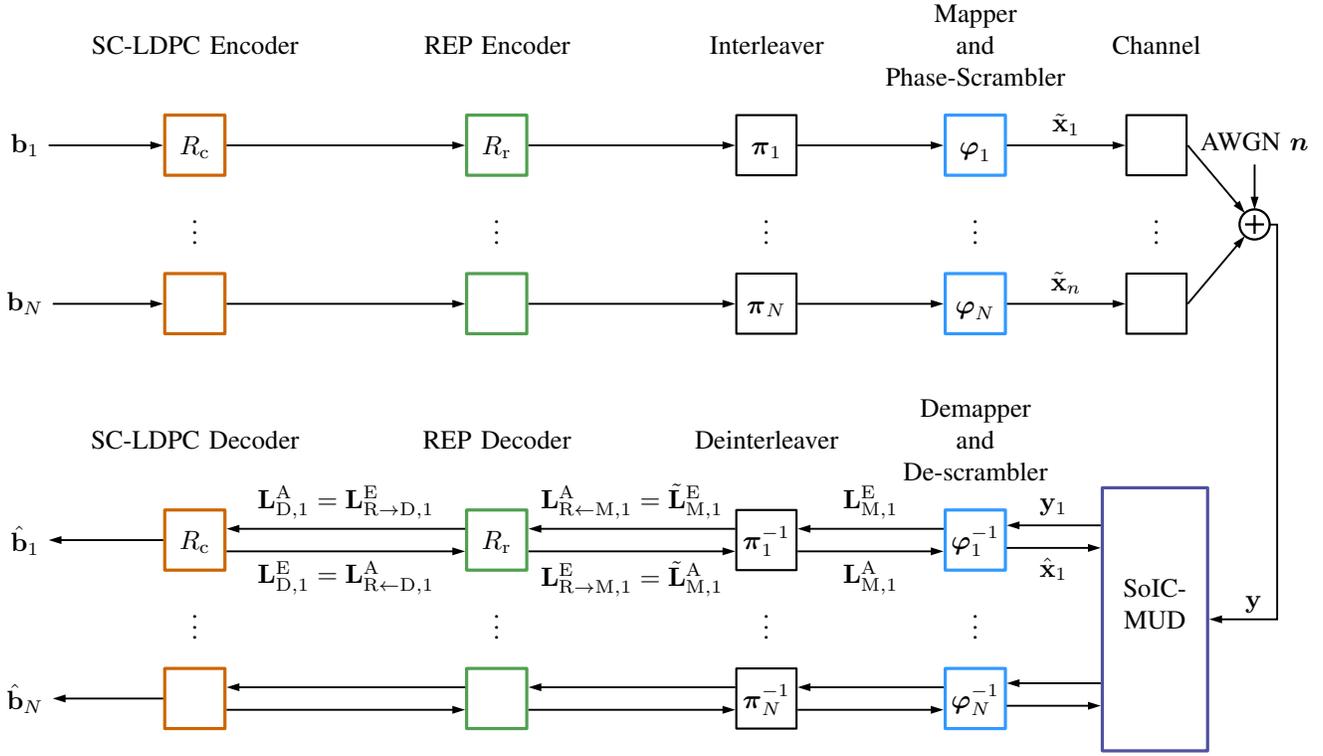
\begin{figure*}
\begin{centering}
\begin{tikzpicture}[scale=0.1]
% Place nodes using a matrix 	
\matrix (TXchains) [row sep=2.5mm, column sep=4mm] { 		
%-------------------------------------------------------------------- 
\node[]  () {};&
\node[]  () {SC-LDPC Encoder};&
\node[text width=4cm, align = center]  () {REP Encoder};&
\node[]  () {Interleaver};&
\node[align=center]  () {Mapper \\ and \\ Phase-Scrambler};&
\node[]  () {Channel};&
&
\\
\node[]  (bitsource1) {$\mathbf{b}_{1}$};&
\node[dspsquare,very thick,draw = myorgange]       (fec1) {$R_{\mathrm{c}}$}; & 		
\node[dspsquare,very thick,draw = mygreen]       (ce1) {$R_{\mathrm{r}}$}; & 		
\node[dspsquare,thick]       (interleaver1) {\,$\boldsymbol{\pi}_{1}$\,}; &		
\node[dspsquare,very thick,draw = lightblue]   (mapper1) {$\boldsymbol{\varphi}_{1}$}; &
\node[dspsquare,thick]   (channel1) {}; &
&
\\
\node[]  () {};&				
\node[]  () {$\vdots$};&
\node[]  () {$\vdots$};&
\node[]  () {$\vdots$};&
\node[]  () {$\vdots$};&
\node[]  () {$\vdots$};&
\node[dspadder]                     (adder) {};&
%\node[align=center, anchor=north, above = 0.5cm of adder] (noise) {};
\\
\node[] ()  (bitsourceN) {$\mathbf{b}_{N}$};& 
\node[dspsquare,very thick,draw = myorgange]       (fecN) {}; & 		
\node[dspsquare,very thick,draw = mygreen]       (ceN) {}; & 		
\node[dspsquare,thick]       (interleaverN) {\,$\boldsymbol{\pi}_{N}$\,}; &		
\node[dspsquare,very thick,draw = lightblue]   (mapperN) {$\boldsymbol{\varphi}_{N}$}; &
\node[dspsquare,thick]   (channelN) {}; &
&				
\\
\node[]  () {};&
\node[] () {};&
\node[] () {};&
\node[]  () {};&
\node[]   () {};&
\node[]  () {};&
\node[]   () {};&
\\
\node[]  () {};&
\node[]  () {SC-LDPC Decoder};&
\node[]   () {REP Decoder};&
\node[]  () {Deinterleaver};&
\node[align=center]   () {Demapper \\ and \\ De-scrambler};&
\node[]   () {};&
\node[]   () {};&
\\
\node[] ()  (bitsinke1) {$\hat{\mathbf{b}}_{1}$};&
\node[dspsquare,very thick,draw = myorgange]   (Noisedecoder1) {$R_{\mathrm{c}}$}; &
\node[dspsquare,very thick,draw = mygreen]   (decoder1) {$R_{\mathrm{r}}$}; &
\node[dspsquare,thick]       (deinterleaver1) {\,$\boldsymbol{\pi}_{1}^{-1}$\,}; &
\node[dspsquare,very thick,draw = lightblue]       (demapper1) {$\boldsymbol{\varphi}^{-1}_{1}$}; &			
\\
\node[]  () {};&
\node[]  () {$\vdots$};&
\node[]  () {$\vdots$};&
\node[]  () {$\vdots$};&
\node[]  (cNode) {$\vdots$};&
&
\\	
\node[]  (bitsinkeN) {$\hat{\mathbf{b}}_{N}$};&
\node[dspsquare,very thick,draw = myorgange]   (NoisedecoderN) {}; &	
\node[dspsquare,very thick,draw = mygreen]   (decoderN) {}; &
\node[dspsquare,thick]       (deinterleaverN) {\,$\boldsymbol{\pi}_{N}^{-1}$\,}; &
\node[dspsquare,very thick,draw = lightblue]       (demapperN) {$\boldsymbol{\varphi}^{-1}_{N}$}; & 			
\\
};		
\node[dspfilter,very thick,right=of cNode,xshift=5mm,yshift=0mm,minimum height=35mm,draw = myblue]                     (MUD) {SoIC- \\ MUD};
%\matrix (RXchains) [row sep=2.5mm, column sep=3mm,xshift=5mm, left=of MUD] { 		
%--------------------------------------------------------------------  		

%};
		
	% Draw connections
	\draw[dspconn] (bitsource1) -- 				(fec1);
    \draw[dspconn] (fec1) -- 				      (ce1);
	\draw[dspconn] (ce1) -- 				(interleaver1);
    \draw[dspconn] (interleaver1) -- 				(mapper1);
    \draw[dspconn] (mapper1) -- 				(channel1) node[midway,above] {$\tilde{\mathbf{x}}_{1}$};
    \draw[dspconn] (channel1.east) -- 				(adder);
	\draw[dspconn] (bitsourceN) -- 				(fecN);
    \draw[dspconn] (fecN) -- 				      (ceN);
	\draw[dspconn] (ceN) -- 				(interleaverN);
    \draw[dspconn] (interleaverN) -- 				(mapperN);
    \draw[dspconn] (mapperN) -- 				(channelN) node[midway,above] {$\tilde{\mathbf{x}}_{n}$};
    \draw[dspconn] (channelN.east) -- 				(adder);
    \draw[dspconn,<-] (adder)  -- +(0,8) node[near end,above, yshift=2mm] {AWGN $\boldsymbol{n}$};
     \draw[dspconn] (adder.east) |- +(1,0) |- (MUD.east) node[near end,right, xshift=-1mm,yshift=2mm] {$\mathbf{y}$};
    %\draw[dspconn] (adder.east) --				($(adder.east)+(1,0)$) -- ($$) -- (MUD.east);
    %\draw[dspconn]  (MUD.west) |- 				(demapper1);
    \draw[dspconn] ($(MUD.west)+(0,12.5)$) -- node[above] {$\mathbf{y}_{1}$} 		+(-12.5,0);
    \draw[dspconn,<-] ($(MUD.west)+(0,9.5)$) -- node[below] {$\hat{\mathbf{x}}_{1}$}		+(-12.5,0);
    %\draw[dspconn] ($(demapper1.east)+(0,-0.2)$) -- 				($(MUD.west)+(0,0.8)$);
    \draw[dspconn] ($(demapper1.west)+(0,1.5)$) -- 				($(deinterleaver1.east)+(0,1.5)$) node[midway,above] {$\mathbf{L}^{\mathrm{E}}_{\mathrm{M},1}$};
    \draw[dspconn] ($(deinterleaver1.east)+(0,-1.5)$) -- 				($(demapper1.west)+(0,-1.5)$) node[midway,below] {$\mathbf{L}^{\mathrm{A}}_{\mathrm{M},1}$};
    \draw[dspconn] ($(deinterleaver1.west)+(0,1.5)$) -- 				($(decoder1.east)+(0,1.5)$) node[midway,above] {$\mathbf{L}^{\mathrm{A}}_{\mathrm{R\leftarrow M},1}=\tilde{\mathbf{L}}^{\mathrm{E}}_{\mathrm{M},1}$};
    \draw[dspconn] ($(decoder1.east)+(0,-1.5)$) -- 				($(deinterleaver1.west)+(0,-1.5)$) node[midway,below] {$\mathbf{L}^{\mathrm{E}}_{\mathrm{R\rightarrow M},1}=\tilde{\mathbf{L}}^{\mathrm{A}}_{\mathrm{M},1}$};
    \draw[dspconn] ($(decoder1.west)+(0,1.5)$) -- 				($(Noisedecoder1.east)+(0,1.5)$) node[midway,above] {$\mathbf{L}^{\mathrm{A}}_{\mathrm{D},1}=\mathbf{L}^{\mathrm{E}}_{\mathrm{R\rightarrow D},1}$};
    \draw[dspconn] ($(Noisedecoder1.east)+(0,-1.5)$) -- 				($(decoder1.west)+(0,-1.5)$) node[midway,below] {$\mathbf{L}^{\mathrm{E}}_{\mathrm{D},1}=\mathbf{L}^{\mathrm{A}}_{\mathrm{R\leftarrow D},1}$};
	\draw[dspconn] (Noisedecoder1) -- 				(bitsinke1);
    \draw[dspconn,<-] ($(MUD.west)+(0,-11.5)$) -- 		+(-12.5,0);
    \draw[dspconn] ($(MUD.west)+(0,-8.5)$) -- 		+(-12.5,0);
    %\draw[dspconn] ($(demapperN.east)+(0,-0.2)$) -- 				($(MUD.west)+(0,-2.0)$);
    \draw[dspconn] ($(demapperN.west)+(0,1.5)$) -- 				($(deinterleaverN.east)+(0,1.5)$);
    \draw[dspconn] ($(deinterleaverN.east)+(0,-1.5)$) -- 				($(demapperN.west)+(0,-1.5)$);
    \draw[dspconn] ($(deinterleaverN.west)+(0,1.5)$) -- 				($(decoderN.east)+(0,1.5)$);
    \draw[dspconn] ($(decoderN.east)+(0,-1.5)$) -- 				($(deinterleaverN.west)+(0,-1.5)$);
    \draw[dspconn] ($(decoderN.west)+(0,1.5)$) -- 				($(NoisedecoderN.east)+(0,1.5)$);
    \draw[dspconn] ($(NoisedecoderN.east)+(0,-1.5)$) -- 				($(decoderN.west)+(0,-1.5)$);
    \draw[dspconn] (NoisedecoderN) -- 				(bitsinkeN);
\end{tikzpicture}
\end{centering}
\caption{IDMA system model; all users have the same coding and modulation scheme;
note that boldface letters denote vectors.\label{fig:IDMA-System-Model} The receiver can be implemented in a windowed version (see Sec.~\ref{sec:windowed_dec} for details), however, for simplicity indices related to windowed decoding are omitted.}
\vspace*{-0.4cm}
\end{figure*}

Fig.~\ref{fig:IDMA-System-Model} shows the IDMA system model with
$N$ uncooperative users. Each user encodes and decodes its data separately
using a channel encoder (SC-LDPC code here) of code rate $R_{c}$ and a common serially
concatenated repetition code of rate $R_{r}=\frac{1}{d_{\mathrm{r}}}$ (see \cite{wang2018idma} for details).
Note that the \ac{SC-LDPC} code and, thus, also the code parameters, e.g., degree profile and coupling width $W$, are the same among all the $N$ users. Thus, the total
code-rate is $R_{\mathrm{tot}}=R_{c}R_{r}$. The interleaver
is, on the contrary, user-specific to allow efficient user separation
at the receiver. After interleaving, the coded bits are mapped to
symbols, e.g., using \ac{BPSK}, and transmitted
over the \ac{GMAC}.

The $m$th
received signal (i.e., the $m$th element of $\mathbf{y}$ in Fig.~\ref{fig:IDMA-System-Model})
of all users can be written as
\begin{equation}
y_{m}={\displaystyle \sum_{i=1}^{N}}\sqrt{P_i}h_{i,m}\cdot\underset{:=\widetilde{x}_{i,m}}{\underbrace{x_{i,m}\cdot e^{j\varphi_{i,m}}}}+n_{m}
\end{equation}
where $m$ is the discrete-time index, $\sqrt{P_i}$ denotes the signal power of the $i$th user, $h_{i,m}$ is the uncorrelated (both over time and among different users) small-scale Rayleigh fading channel coefficient, $n_{m}$ is circularly symmetric
(complex-valued) AWGN with zero mean and variance $\sigma_{n}^{2}$,
and $\varphi_{i,m}$ is a pseudo random phase scrambling to avoid ambiguity
of the super-constellation (Cartesian product of all user constellations).
This random phase shift could also be the consequence of, e.g., the
channel and/or explicit ``scrambling'' and we include this into
each user's mapper (only in AWGN channels; for Rayleigh channels this step can be omitted). Throughout this paper, the phases $\varphi_{i,m}$
are independently and uniformly distributed in $\left[0,\pi\right)$.
The output of the mapper of the $i$th user with BPSK modulation at
the $m$th time instant is $\tilde{x}_{i,m}\in\left\{ \pm e^{j\varphi_{i,m}}\right\} $. The “phase scrambling” can improve the superimposed multiuser codeword distance \cite{GSongIT16}, particularly in AWGN channels. The so-called multi-user \ac{SNR} is defined as 
\begin{equation}
\gamma= \frac{\sum_{i=1}^{N} P_i }{\sigma_{n}^{2}}.
\end{equation}

The received signal is first processed by a multi-user detector (MUD).
An optimum MUD is to compute the maximum a posteriori (MAP) probability
of each bit. This requires a complexity of $O\left(M^{N}\right)$
where $M$ denotes the number of constellation symbols per user. The
exponentially increasing complexity with the number of users $N$
prohibits its practical implementation for a large number of users.
Therefore, a sub-optimal soft interference cancellation (SoIC) based
low complexity MUD was proposed in \cite{LiPingIDMACL04}. The sub-optimal
MUD first cancels out the other users' signals; for instance, the $i$th
user's signal is estimated by the conditional minimum mean-square error (MMSE) estimator for BPSK 
\[
\hat{x}_{i}=\mathrm{tanh}\left(\frac{L_{\mathrm{M,}i}^{\mathrm{A}}}{2}\right)\cdot e^{j\varphi_{i}}
\]
based on, e.g., the a posteriori knowledge of the channel decoder
$L_{\mathrm{M,}i}^{\mathrm{A}}$ (the \ac{SC-LDPC} decoder output is $L_{\mathrm{R\leftarrow D,}i}^{\mathrm{A}}$; it is then re-encoded
by a \ac{REP} and re-interleaved with the outputs denoted by $\tilde{L}_{\mathrm{M,}i}^{\mathrm{A}}$
and $L_{\mathrm{M,}i}^{\mathrm{A}}$, respectively). For an arbitrary
user $j$ (the symbol index $m$ is dropped for brevity), the output
of the MUD after the SoIC can be written as
\begin{align*}
y_{j}	=\sqrt{P_{j}}h_{j}\widetilde{x}_{j}+{\displaystyle \sum_{i=1,i\ne j}^{N}\sqrt{P_{i}}h_{i}\left(\tilde{x}_{i}-\hat{x}_{i}\right)}+n.
\end{align*}

Then, each user starts its single user detection and decoding in parallel.
The (soft) demapper computes the log-likelihood-ratio (LLR) of each
bit while treating the residual interference as noise. For BPSK, an
approximation of the true a posteriori LLR can be computed according
to 
\[
L_{\mathrm{M},j}^{\mathrm{E}}=4\sqrt{P_{j}}\frac{\mathrm{Re}\left\{ y_{j}\cdot h_{j}^{*}\cdot e^{-j\varphi_{j}}\right\} }{\sigma_{\mathrm{I},j}^{2}+\sigma_{n}^{2}}
\]
where the noise variance $\sigma_{n}^{2}$,  the random phase shifts
$\varphi_{j}$ and the channel coefficients $h_{j}$  are assumed to be known at the receiver. The interference
power can be estimated by 
\begin{align}
\begin{split}
\sigma_{\mathrm{I},j}^{2}=\mathrm{E}\left[{\displaystyle \left|e^{-j\varphi_{j}}\sum_{i\ne j}\sqrt{P_{i}}h_{i}\left(\tilde{x}_{i}-\hat{x}_{i}\right)\right|^{2}}\right] \\
	=\sum_{i\ne j}P_{i}\left|h_{i}\right|^{2}\left(1-\mathrm{E}\left[\mathrm{tanh}\left(\frac{L_{\mathrm{M,}i}^{\mathrm{E}}}{2}\right)\right]^{2}\right)
\end{split}
\end{align}
where the interference term is assumed to be Gaussian distributed,
provided that the number of users $N$ is large enough and the transmitted
symbols are independent among users (central limit theorem). 

Then, the LLRs are deinterleaved (denoted by $\tilde{L}_{\mathrm{M},j}^{\mathrm{E}}=L_{\mathrm{R}\leftarrow M,j}^{\mathrm{A}}$
which means the extrinsic message from the MUD corresponds to the a priori
knowledge of the REP obtained by the MUD) and sent to a repetition decoder.
The extrinsic message from the repetition code to the LDPC decoder
is given by 
\[
L_{\mathrm{D},j,m}^{\mathrm{A}}=L_{\mathrm{R\rightarrow D},j,m}^{\mathrm{E}}=\sum_{k=md_{\mathrm{r}}}^{\left(m+1\right)d_{\mathrm{r}}-1}\tilde{L}_{\mathrm{M,}j,k}^{\mathrm{E}}.
\]
Subsequently, channel decoding can be performed by
the corresponding channel decoder.

\iffalse
The SoIC-based MUD is quite simple and requires only few multiplications
and additions for variance estimation, interference subtraction and
LLR computation (see \cite{LipingIDMATWC06}). The complexity becomes
$O\left(MN\right)$ and many computations can be carried out in parallel
when compared to SIC. Particularly, when all the users apply an LDPC
code with the same parity check $\mathbf{H}$-matrix, the coordination
and code design among users can be dramatically reduced.
\fi

\section{Spatially Coupled LDPC}
%\begin{itemize}
%\item Basics
%\item Density evolution
%\end{itemize}
%
%\begin{itemize}
%\item universality
%\item DE thresholds
%\item universality, visualize via EXIT (maybe Rayleigh)
%\end{itemize}

We follow the definitions in \cite{protograph_mitchell}, i.e., consider \ac{SC-LDPC} code ensembles defined by their protograph matrix $\Bm$.
Protographs can be seen as a blueprint of larger graphs, where $\mathit{S}$ copies of the protograph are randomly connected by edge permutations. Each non-zero entry of the corresponding base matrix $\Bm$ represents the number of connected edges to this node type.
For further details we refer interested readers to \cite{protograph_mitchell, MCT_URBANKE}.

For the sake of spatial coupling, $\mathbf{B}$ can be divided into $W$ sub-matrices  $\mathbf{B_{\mathit{i}}}$ of dimension $\mathit{M'\times N'}$ \cite{protograph_mitchell}, i.e.,
$$\mathbf{B}=\left[\begin{array}{c}
\mathbf{B_{\mathit{0}}}\\
\vdots\\
\mathbf{B}_{\mathit{W-1}}
\end{array}\right]_{WM'\times N'}.$$
The approach from \cite{protograph_mitchell} is used to construct the \ac{SC-LDPC} protograph matrix $\mathbf{B_{L,W}}$, where $\mathit{L}$ denotes the replication factor (i.e., the number of sub-blocks). For a terminated code and a coupling window $\mathit{W=\mathrm{3}}$, we get
$$\mathbf{B}_{L,W=\mathrm{3}}=\left[\begin{array}{cccc}
\mathbf{B}_{0} & \phantom{\ddots_{1}} & \phantom{\ddots_{1}} & \phantom{\ddots_{1}}\\
\mathbf{B}_{1} & \mathbf{B}_{0} & \phantom{\ddots_{1}} & \phantom{\ddots_{1}}\\
\mathbf{B}_{2} & \mathbf{B}_{1} & \ddots & \phantom{\ddots_{1}}\\
\phantom{\ddots_{1}} & \mathbf{B}_{2} & \ddots & \mathbf{B}_{0}\\
\phantom{\ddots_{1}} & \phantom{\ddots_{1}} & \ddots & \mathbf{B}_{1}\\
\phantom{\ddots_{1}} & \phantom{\ddots_{1}} & \phantom{\ddots_{1}} & \mathbf{B}_{2}
\end{array}\right]_{\left(L+W-1\right)M'\times LN'}.$$

Finally, a \emph{lifting} step with lifting factor $Z$ results in the parity-check matrix $\Hm_{L,W,Z}$.

\subsection{Density Evolution for the \ac{GMAC} with iterative detection/decoding}
\label{sec:de}
The decoding threshold can be obtained via density evolution \cite{MCT_URBANKE, Chung_SPA_GA}. We apply a \ac{GA}, i.e., we only track the mean value $\mu$ of messages passed along within the decoder and the iterative \ac{MUD} with transfer function $f_{\mathrm{MUD}}(\mu_{\mathrm{A,MUD}})$ as in \cite{Schmalentenbrink}.

We denote the entry of $\Bm_{L,W}$ in the $j$-th row and the $i$-th column as $B_{j,i}$.
Let $\mathit{\mu_{i\leftarrow j}}$ denote the mean value of messages passed from \ac{CND} $\mathit{c_{j}}$ with spatial position $j$ to a connected \ac{VND} $\mathit{v_{i}}$ at spatial position $i$ and let $\mu_{i\rightarrow j}$ denote the mean value passed from \ac{VND} $\mathit{v_{i}}$ to \ac{CND} $\mathit{c_{j}}$. The update rules become\footnote{For readability, we only consider edges where $B_{j,i}\neq 0$, all unconnected edges ($B_{j,i}=0$) virtually transmit $\mu=0$.} \cite{Schmalentenbrink, Chung_SPA_GA}
 
\begin{align}
 \mu_{i\leftarrow j}=\phi^{-1} 
\Biggl( 1-  &  \left[1-\phi\left( \mu_{i\rightarrow j}\right)\right]^{B_{j,i}-1} 
\\
&  \cdot \prod_{k=1;k\neq i}^{LN'}\left[1-\phi\left(\mu_{i\rightarrow j}\right)\right]^{B_{j,k}}\Biggr) \nonumber
\end{align}

with $\phi\left(\mu\right)$ as in \cite{Chung_SPA_GA}
$$\phi\left(x\right)=\begin{cases}
1-\frac{1}{\sqrt{4\pi x}}\int_{-\infty}^{\infty}\tanh\left(\frac{u}{2}\right)\exp\left(-\frac{\left(u-x\right)^{2}}{4x}\right)\mathrm{d\mathit{u},} & x>0\\
1, & x=0
\end{cases}$$

Due to the serial concatenation of a \ac{REP} of rate $R_{r}=\frac{1}{d_r}$, the mean of the messages passed from VND $v_i$ (including the REP code) to the MUD is therefore 
\begin{equation}
\mu_{D\leftarrow i}=\left(d_{r}-1\right)\mu_{D\rightarrow i}+ \sum_{k=1}^{LM'}B_{k,i}\cdot\mu_{i\leftarrow k}.\label{eq:mu_di}
\end{equation}
Here $\mu_{D\rightarrow i}$ denotes the mean of the message from \ac{MUD} to \ac{VND} $v_i$ after the PIC processing at MUD nodes. These updated messages can be written as \cite{wang2018idma} 
$$\mu_{D\rightarrow i}=\frac{4}{N\sigma_{n}^{2}+\left(N-1\right)\cdot\phi\left(\mu_{_{D\leftarrow i}}\right)}.$$

The variable node update from \ac{VND} $\mathit{v_{i}}$ to \ac{CND} $\mathit{c_{j}}$ is $$\mu_{i\rightarrow j}=d_{r}\mu_{D\rightarrow i}+\left(B_{i,j}-1\right)\cdot\mu_{i\leftarrow j}+\sum_{k=1,k\neq j}^{LM'}B_{k,i}\cdot\mu_{i\leftarrow k}.$$

%old
%We use the approximation \cite{PhDSchreck}  $J(\mu)\approx\left(1-2^{-H_{1}\cdot\left(2\mu\right)^{H_{2}}}\right)^{H_{3}}$ and $J^{-1}(I)\approx\frac{1}{2}\left(-\frac{1}{H_{1}}\cdot\log_{\mathrm{2}}\left(1-I^{\frac{1}{H_{3}}}\right)\right)^{\frac{1}{H_{2}}}$ with $\mathit{H_{\mathrm{1}}=\mathrm{0.3073}}$, $\mathit{H_{\mathrm{2}}=\mathrm{0.8935}}$ and $\mathit{H_{\mathrm{3}}=\mathrm{1.1064}}$.

In this work, we use codes as proposed in \cite{protograph_mitchell}:
\begin{itemize}
\item $C_1$: \ac{SC-LDPC} $(d_v=3,d_c=6,L,W=3)$ code with ${R_{c,L \to \infty}=0.5}$ and $\Bm_0=\Bm_1=\Bm_2=[1\quad1]$ 
\item $C_2$: \ac{SC-LDPC} $(d_v=3,d_c=4,L,W=2)$ code with ${R_{c,L \to \infty}=0.25}$, $$\Bm_0 =\begin{bmatrix}
    1 & 1 & 0 & 0 \\
    0 & 1 & 1 & 0 \\
    0 & 0 & 1 & 1
  \end{bmatrix} \quad \textrm{and} \quad \Bm_1 =\begin{bmatrix}
    0 & 0 & 1 & 1 \\
    1 & 0 & 0 & 1 \\
    1 & 1 & 0 & 0
  \end{bmatrix}.$$
\end{itemize}

Table~\ref{tab:thresholds} shows the decoding thresholds $\gamma_{\mathrm{un}}^*$ and $\gamma_{\mathrm{sc}}^*$ for the uncoupled and the SC-LDPC ensemble, respectively.
As expected a higher node degree degrades the uncoupled thresholds, however, the \ac{SC-LDPC} codes show an improved threshold which coincides well with the effect of \emph{threshold saturation} \cite{Coupl11BMS}.

\begin{table}

\caption{Density evolution-based decoding thresholds of different codes for 8 users over the \ac{GMAC} and $d_r$ such that $R_{sum}=1$ and the Shannon limit $\gamma_{\mathrm{Sh}}=0~\operatorname{dB}$ \label{tab:decoding_thresholds}}

\vspace*{-0.4em}
\centering
\begin{tabular}{ l c c c r }
\label{tab:thresholds}
$d_r$ & $d_v$ & $d_c$ & uncoupled ($\gamma_{\mathrm{un}}^{*}$) & coupled\tablefootnote{Remark: for a better comparison the rate loss due to termination effects is not considered here.} ($\gamma_{\mathrm{SC}}^{*}$)\\
\hline 
\hline  
 4 & 3 & 6 & $2.54 \operatorname{dB}$ & $1.55 \operatorname{dB}$ \\
 4 & 4 & 8 &  $3.43 \operatorname{dB}$ & $1.42 \operatorname{dB}$ \\
 4 &  5 & 10 & $4.11 \operatorname{dB}$ & $1.33 \operatorname{dB}$ \\
 4 & 6 & 12 & $4.62 \operatorname{dB}$ & $1.13 \operatorname{dB}$ \\
  \hline
 2 & 3 & 4 & $3.96 \operatorname{dB}$ & $0.74 \operatorname{dB}$ \\
 2 & 6 & 8 & $14.98 \operatorname{dB}$ & $0.69 \operatorname{dB}$ \\
 2 &  9 & 12 & -- & $0.69 \operatorname{dB}$ \\
\end{tabular}

\end{table}

\subsection{EXIT analysis}

\begin{figure}[t]
	\centering
	\resizebox{0.9\columnwidth}{!}{
	\begin{tikzpicture} [spy using outlines=
	{magnification=3, connect spies}]
\begin{axis}[
width=\linewidth,
height=\linewidth,
xmajorgrids,
yminorticks=true,
ymajorgrids,
yminorgrids,
legend pos=south east,        
xlabel={$I_{A,V},I_{E,C}$},
ylabel={$I_{E,V},I_{A,C}$},
%ymode=lin,
%xmode=lin,
mark size=1.5pt,
xmin=0,
xmax=1,
ymin=0,
ymax=1
]	
% \pgfplotstableread \datatable

\addplot[color= mittelblau, ultra thick,each nth point={10}] table [x index=0,y index=1] {data/exit_4.dat};
\addlegendentry{MUD+REP+VND}

\addplot[color= rot, ultra thick,each nth point={10}] table [x index=0,y index=1] {data/exit_3.dat};
\addlegendentry{CND}

%\addplot[color= yellow, thick,each nth point={2}] table [x index=0,y index=1] {data/exit_2.dat};
%\addlegendentry{Demapper $d_{c}=3$}

\addplot[color= apfelgruen, ultra thick,each nth point={1}] table [x index=0,y index=1] {data/exit1_1.dat};
\addlegendentry{SC-LDPC Trajectory}

\addplot[color= orange, densely dashed, ultra thick,each nth point={1}] table [x index=0,y index=1] {data/exit_1.dat};
\addlegendentry{Block LDPC Trajectory}

\coordinate (spypoint1) at (axis cs:0.12,0.55);
\coordinate (magnifyglass1) at (axis cs:0.6,0.49);

\node[align=center,color=orange] (node1) at (axis cs:0.1585,0.5695) {\tiny{Block LDPC}};
\node[align=center,color=orange] (node1) at (axis cs:0.144,0.549) {\tiny{gets stuck}};

\end{axis}

\spy [black,width=4.5cm,height=2.4cm] on (spypoint1) in node [fill=none] at (magnifyglass1);

\end{tikzpicture}}
	%\vspace*{-0.8cm}
	\caption{\ac{EXIT} chart of a $(3,6)$-\ac{SC-LDPC} \ac{IDMA} system at \ac{SNR} $\gamma=2.3 \operatorname{dB}$ and $d_r=4$ and 8 user, i.e., $R_{sum}=1$. }
	\label{pgf:EXIT}		
	\vspace*{-0.5cm}
\end{figure}
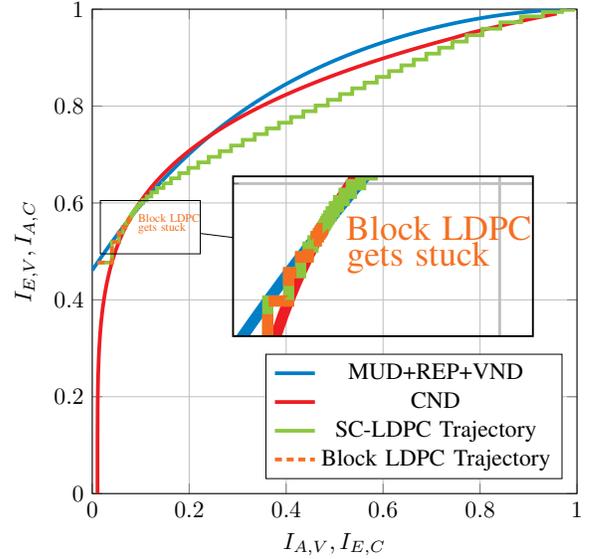

Fig.~\ref{pgf:EXIT} shows the \ac{EXIT} chart for the proposed \ac{MUD} system with a \ac{SC-LDPC} code at an \ac{SNR} of $\gamma=2.3 \operatorname{dB}$, i.e., below the uncoupled decoding threshold of $\gamma_\mathrm{un}^*=2.54 \operatorname{dB}$. As it can be seen, convergence is possible although no open \emph{decoding tunnel} exists in the \ac{EXIT} chart (cf. \emph{micro-convergence} in \cite{Schmalentenbrink}).
As successful decoding is possible even if \ac{VND} and \ac{CND} curves intersect, no extensive degree profile matching of the code components is required anymore. This intuitively visualizes the \emph{universal} behavior of \ac{SC-LDPC} codes as their decoding thresholds are (approximately) universal for varying channel conditions \cite{Schmalentenbrink,cammerer2016wave}.

\section{Windowed MUD and channel decoder}
\label{sec:windowed_dec}

So far, we considered \emph{full} \ac{BP} decoding, meaning that all spatial positions are updated in parallel for each iteration. However, the authors in \cite{iyengar2013windowed} showed that a windowed decoding scheme reduces decoding complexity without significant \ac{BER} performance loss. Throughout this work, we assume a constant number of iterations per window shift, yet, an adaptive implementation is possible \cite{klaiber2018windowed}.

\subsection{Windowed receiver}

The windowed decoder makes use of the fact that the \ac{BER} per spatial position converges in a wavelike manner, i.e., subsequent blocks can only be decoded if the previous blocks have been successfully decoded. Therefore, it is sufficient to only update nodes within a few spatial positions (\emph{active} window), i.e., per decoding iteration only $W_d\ge W$ sub-blocks are active. We keep the decoding structure as in Fig.~\ref{fig:IDMA-System-Model}, but activate the same window for all users in parallel (i.e., all nodes at the same spatial positions). The windowed \ac{MUD} estimates the \ac{SNR} per sub-block (see Sec.~\ref{sec:interleaver} for the intuition behind) and updates its outgoing messages accordingly. Finally, the window position is shifted by one spatial position after $I_{max}$ decoding iterations until all sub-blocks are decoded. For further details on the initialization, see \cite{klaiber2018windowed}.

\subsection{Subblock interleaving}
\label{sec:interleaver}
The general idea of \ac{IDMA} systems is to separate users by user-specific interleavers which are straightforward to implement for block-codes (besides complexity considerations).
However, for \ac{SC-LDPC} codes (and also due to the windowed receiver scheme) these interleavers require some further attention regarding the interleaving depth.  

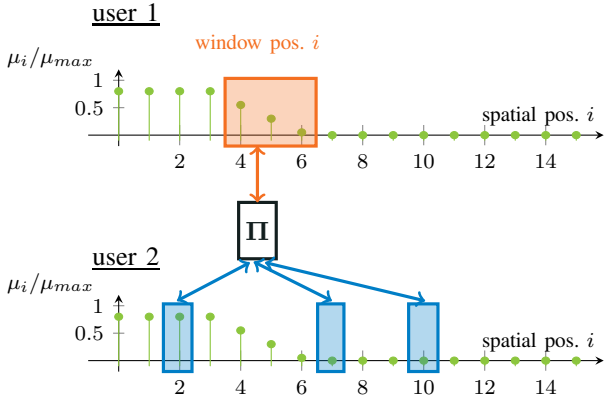
\begin{figure}
	\centering
	\begin{tikzpicture}

\def\scale{1};
\def\yscale{1.8};
\def\xofset{0.25};

\begin{axis}[
axis lines=middle,
axis equal image,
%y dir=reverse,
yscale=\yscale,scale=\scale,
xmin=-1,xmax=15.9,
ymin=-0.2,ymax=1.2,
xlabel={spatial pos. $i$},ylabel={$\mu_i/ \mu_{max}$},
ylabel style ={font=\footnotesize,anchor=east,xshift=-0.2cm,yshift=0.15cm}, x label style ={font=\footnotesize},
yticklabel style = {font=\footnotesize},
xticklabel style = {font=\footnotesize}
]

\addplot [ycomb, apfelgruen,fill=apfelgruen,mark=*,mark size=1.5,yscale=0.5] coordinates { (0,1.8) (1,1.8) (2,1.8) (3,1.8) (4,1.3) (5,0.8) (6,0.3) (7,0.2) (8,0.2) (9,0.2) (10,0.2) (11,0.2) (12,0.2) (13,0.2) (14,0.2) (15,0.2)};

\end{axis}

\draw[very thick,color=orange,fill=orange,fill opacity=0.3] (1.82,0.0) rectangle (3.03,0.9);

\begin{axis}[
at={(0cm,-3cm)},
axis lines=middle,
axis equal image,
%y dir=reverse,
yscale=\yscale,scale=\scale,
xmin=-1,xmax=15.9,
ymin=-0.2,ymax=1.2,
xlabel={spatial pos. $i$},ylabel={$\mu_i/ \mu_{max}$},
ylabel style ={font=\footnotesize,anchor=east,xshift=-0.2cm,yshift=0.15cm},x label style ={font=\footnotesize},
yticklabel style = {font=\footnotesize},
xticklabel style = {font=\footnotesize}
]
\addplot [ycomb, apfelgruen,fill=apfelgruen,mark=*,mark size=1.5,yscale=0.5] coordinates { (0,1.8) (1,1.8) (2,1.8) (3,1.8) (4,1.3) (5,0.8) (6,0.3) (7,0.2) (8,0.2) (9,0.2) (10,0.2) (11,0.2) (12,0.2) (13,0.2) (14,0.2) (15,0.2)};

%\draw[very thick,color=apfelgruen,fill=apfelgruen,fill opacity=0.3] (25,0.0) rectangle (34.5,1.25);
%\draw[very thick,color=mittelblau,fill=mittelblau,fill opacity=0.3] (76,0.0) rectangle (84.5,1.25);
%\draw[very thick,color=rot,fill=rot,fill opacity=0.3] (106,0.0) rectangle (115,1.25);

\end{axis}

\draw[very thick,color=mittelblau,fill=mittelblau,fill opacity=0.3] (1.0,-2.1) rectangle (1.38,-3);
\draw[very thick,color=mittelblau,fill=mittelblau,fill opacity=0.3] (3.05,-2.1) rectangle (3.4,-3);
\draw[very thick,color=mittelblau,fill=mittelblau,fill opacity=0.3] (4.26,-2.1) rectangle (4.65,-3);

\node[thick] (user) at (0.5,1.75) {\underline{user 1}};
\node[thick] (user) at (0.5,-1.5) {\underline{user 2}};
\node[thick,orange] (user) at (2.25,1.35) {\footnotesize{window pos. $i$}};

\draw[very thick,color=anthrazit] (1.75+\xofset,-1.5) rectangle (2.25+\xofset,-0.75)node[pos=.5] {$\mathbf{\Pi}$} ;

\draw[very thick,color=orange,<->] (2.25,-0.75) --(2.25,-0);
\draw[very thick,color=mittelblau,<->] (2.15,-1.55) --(1.18,-2.06);
\draw[very thick,color=mittelblau,<->] (2.2,-1.54) --(3.24,-2.06);
\draw[very thick,color=mittelblau,<->] (2.35,-1.55) --(4.48,-2.06);

\end{tikzpicture}
	%\vspace*{-0.8cm}
	\caption{Illustration of wavelike-decoding in a two-user \ac{IDMA} system with a \emph{full} interleaver accessing random spatial positions of user 2 that are not yet converged.}
	\label{fig:int_pattern}		
	\vspace*{-0.1cm}
\end{figure}

In (\ref{eq:mu_di}), we assumed that the incoming messages from the \ac{MUD} are all from spatial position $i$ for all users, i.e., they have the same underlying statistics and, thus, $\mu_{D\leftarrow i}$ is the same for all users. However, when assuming random interleaving over the whole $\Bm_{SC}$, the update in (\ref{eq:mu_di}) changes to
$$\tilde{\mu}_{D} = \tilde{\mu}_{D\leftarrow i}=\frac{1}{LN'} \sum_{k=1}^{LN'}\mu_{D\leftarrow k}$$
and, thus, a \emph{full} interleaver potentially destroys the locality of the \ac{SC-LDPC} code and spatial positions with high interference noise (yet \emph{unconverged} positions) hinder \emph{wavelike convergence}. Further, the windowed \ac{MUD} does not even update spatial positions outside the currently active window. 
Intuitively, this can be explained as illustrated in Fig.~\ref{fig:int_pattern} for the two-user scenario. As the current decoding progress of spatial position $i$ shares messages with the other users, a random interleaver causes the access of random spatial positions (messages potentially not yet updated) of the other users. This means a user observes the high interference noise as if no (only little) \emph{a priori} knowledge at the receiver exists. Thus, the overall performance can be approximated by a non-iterative scheme, where the initial estimate of the \ac{MUD} provides a lower bound on its performance, i.e., $\tilde{f}_{\mathrm{MUD}}(\mu_{\mathrm{A,MUD}}) \approx f_{\mathrm{MUD}}(0)$.

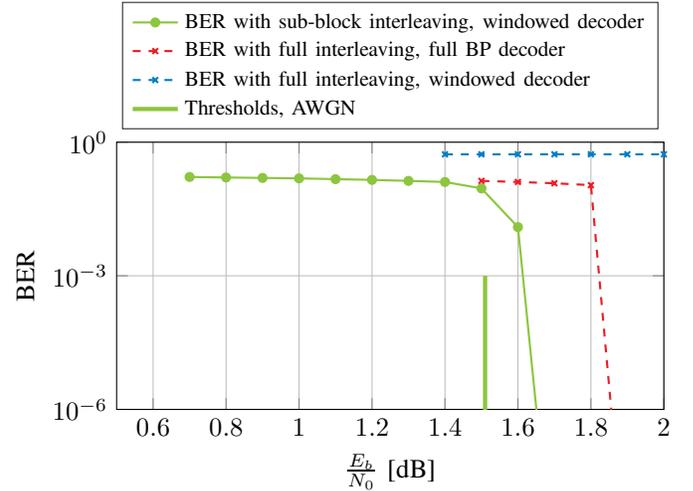
\begin{figure}
	\centering
	\begin{tikzpicture}

\pgfplotsset{compat=1.12}
\begin{axis}[
        width=\linewidth,
	    height=0.58\linewidth,
        xmajorgrids,
        yminorticks=true,
        ymajorgrids,
        yminorgrids,
        legend pos=south west,        
        legend style={at={(0.5,1.525)},
      anchor=north, legend columns=1, legend cell align=left,align=left,draw=white!15!black, font=\footnotesize},
        xlabel={$\frac{E_{b}}{N_0}$ [dB]},
        ylabel={BER},
        ymode=log,
        mark size=1.5pt,
        xmin=0.5,
        xmax=2.0,
        ymin=1e-6,
        ymax=1
    ]	

	\addplot[mark=*,color=apfelgruen, thick] table {data/ber4.dat};
	\addlegendentry{BER with sub-block interleaving, windowed decoder}

	\addplot[mark=x,color=rot, dashed, thick] table {data/BER_fullInt1.dat};
	\addlegendentry{BER with full interleaving, full BP decoder}

	\addplot[mark=x,color=mittelblau, dashed, thick] table {data/BER_fullInt_Win1.dat};
	\addlegendentry{BER with full interleaving, windowed decoder}

      \addplot[color= apfelgruen, ultra thick] coordinates {(1.51, 1e-10) (1.51,1e-3)};
      \addlegendentry{Thresholds, AWGN}

\end{axis}

\end{tikzpicture}
	\vspace*{-0.8cm}
	\caption{\ac{BER} performance of the \ac{SC-LDPC} $(d_v=3,d_c=6)$ \ac{IDMA} system with 8 users and $d_r=2$ for different interleaver implementations.}
	\label{pgf:BER_int}		
	\vspace*{-0.2cm}
\end{figure}

To keep the spatial structure of the code, we propose to use sub-block interleavers which only permute locally within a spatial position. This suffices for the required user separation, but still maintains the locality of the \ac{SC-LDPC} code. However, the price to pay is a potentially smaller interleaver size (or larger sub-block size) and, thus, a slightly degraded \ac{BER} performance of the \ac{IDMA} system. Fig.~\ref{pgf:BER_int} compares the BER performance for a system with the proposed sub-block interleaver (simulation parameters in Sec.~\ref{sec:results}). It can be seen that the decoding performance under full interleaving together with windowed decoding degrades to the expected non-iterative performance, i.e., each user sees the SINR $\frac{1}{\left(N-1\right)+N\sigma_{n}^{2}}$ while the sub-block interleaving yields results close to $\gamma_{\mathrm{SC}}^{*}$ as provided in Table~\ref{tab:thresholds}. If full BP decoding is used with full interleaving, the performance can be dramatically enhanced with 1500 allowed iterations, but is still worse than the sub-block interleaved version with windowed decoding.

%\begin{itemize}
%\item *we should explain the problem (wavelike decoding here)*
%\item @Xiaojie: *permutations need to be checked is it ok if $\mathbf{\Pi}_0^i =\mathbf{\Pi}_1^i$*
%\item *per row or per column constant interleaving?*
%\item *$\Hm_0$ not yet defined*
%\end{itemize}

%For a terminated code and a coupling window $\mathit{W=\mathrm{3}}$, we get
%$$\mathbf{H}_{L,W=\mathrm{3}}=\left[\begin{array}{cccc}
%\mathbf{\Pi}_0^i\mathbf{H}_{0} & \phantom{\ddots_{1}} & \phantom{\ddots_{1}} & \phantom{\ddots_{1}}\\
%\mathbf{\Pi}_0^i \mathbf{H}_{1} &\mathbf{\Pi}_1^i \mathbf{H}_{0} & \phantom{\ddots_{1}} & \phantom{\ddots_{1}}\\
%\mathbf{\Pi}_0^i \mathbf{H}_{2} & \mathbf{\Pi}_1^i \mathbf{H}_{1} & \ddots & \phantom{\ddots_{1}}\\
%\phantom{\ddots_{1}} & \mathbf{\Pi}_1^i \mathbf{H}_{2} & \ddots & \mathbf{\Pi}_L^i \mathbf{H}_{0}\\
%\phantom{\ddots_{1}} & \phantom{\ddots_{1}} & \ddots & \mathbf{\Pi}_L^i \mathbf{H}_{1}\\
%\phantom{\ddots_{1}} & \phantom{\ddots_{1}} & \phantom{\ddots_{1}} & \mathbf{\Pi}_L^i \mathbf{H}_{2}
%\end{array}\right]_{\left(L+W-1\right)M'\times LN'}.$$

\section{Simulation results}\label{sec:results}
%\begin{itemize}
%\item AWGN, compare to prev. work \textcolor{green}{Xiaojie}
%\item BER vs. thresholds
%\item add plot ``universality'' (gap to cap comparison with block codes)
%\item Rayleigh (only BER)
%\end{itemize}
We consider an \ac{IDMA} system with $N=8$ users and a targeted sum-rate of $R_{\mathrm{sum}}=N\frac{R_{c}}{d_{r}}=1$. Both, AWGN and Rayleigh fading channel models are considered. For the AWGN channel case, an explicit phase scrambling is included into the symbol-mapper to improve the distance between multi-user superimposed codewords \cite{GSongIT16}. For Rayleigh fading channels, we consider that the channel states are uncorrelated among users and vary rapidly from symbol to symbol (ergodic fading, i.e., uncorrelated fast fading) and, therefore, a phase scrambler is not necessary in Rayleigh channels. Furthermore, the received signal power levels are assumed to be the same, i.e., $P_{i}=\frac{1}{N}, \forall i$ and \ac{BPSK} is used as modulation format. For the unequal-power case, the repetition code can be used as ''power equalizer'' (see \cite{wang2018idma}).
 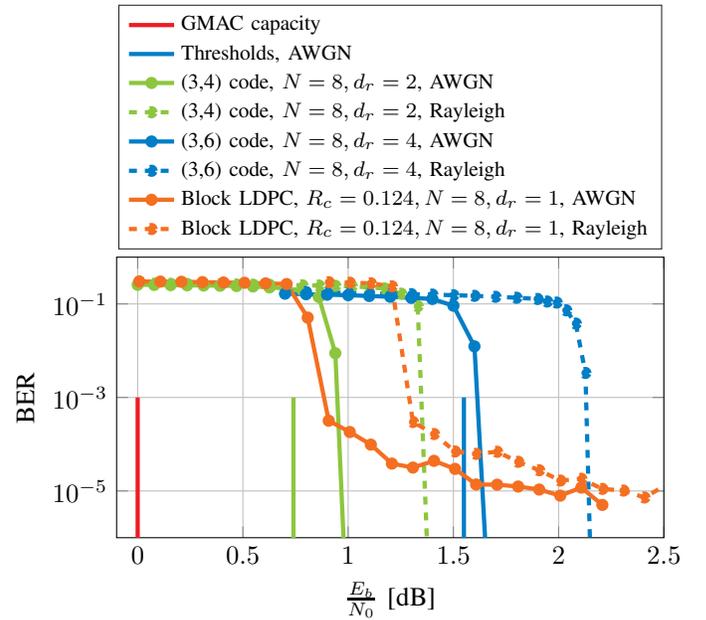
\begin{figure}
	\centering
	\begin{tikzpicture}

\pgfplotsset{compat=1.12}
\begin{axis}[
        width=\linewidth,
	    height=0.6\linewidth,
        xmajorgrids,
        yminorticks=true,
        ymajorgrids,
        yminorgrids,
        legend pos=south west,        
        legend style={at={(0.5,1.9)},
      anchor=north, legend columns=1, legend cell align=left,align=left,draw=white!15!black, font=\footnotesize},
        xlabel={$\frac{E_{b}}{N_0}$ [dB]},
        ylabel={BER},
        ymode=log,
        mark size=1.5pt,
        xmin=-0.1,
        xmax=2.5,
        ymin=1e-6,
        ymax=1
    ]	
\addplot[color= rot, ultra thick] coordinates {(0.0, 1e-10) (0.0,1e-3)};
\addlegendentry{GMAC capacity}

\addplot[color= mittelblau, ultra thick] coordinates {(1.55, 1e-10) (1.55,1e-3)};
\addlegendentry{Thresholds, AWGN}

	\addplot[mark=*,color=apfelgruen, ultra thick] table {data/ber2.dat};
	\addlegendentry{(3,4) code, $N=8, d_{r}=2$, AWGN}

	\addplot[mark=*,color=apfelgruen, ultra thick,dashed] table {data/ber1.dat};
	\addlegendentry{(3,4) code, $N=8, d_{r}=2$, Rayleigh}	
	\addplot[mark=*,color=mittelblau, ultra thick] table {data/ber4.dat};
	\addlegendentry{(3,6) code, $N=8, d_{r}=4$, AWGN}
      \addplot[mark=*,color=mittelblau, ultra thick,dashed] table {data/ber3.dat};
	\addlegendentry{(3,6) code, $N=8, d_{r}=4$, Rayleigh}
	\addplot[mark=*,color=orange, ultra thick] table {data/BER_LDPC_AWGN.dat};
	\addlegendentry{Block LDPC, $R_{c}=0.124, N=8, d_{r}=1$, AWGN}
	\addplot[mark=*,color=orange, ultra thick,dashed] table {data/BER_LDPC_Ray.dat};
	\addlegendentry{Block LDPC, $R_{c}=0.124, N=8, d_{r}=1$, Rayleigh}

\addplot[color= apfelgruen,ultra thick] coordinates {(0.74, 1e-10) (0.74,1e-3)};	
 
\end{axis}

\end{tikzpicture}
	\vspace*{-0.8cm}
	\caption{BER performance of SC-LDPC-coded IDMA systems with window decoder in AWGN and Rayleigh fading channels; $N=8$ users with equal-power and equal-rate are considered.}
	\label{pgf:BER}		
	\vspace*{-0.5cm}
\end{figure}

Fig.~\ref{pgf:BER} shows the \ac{BER} results for such \ac{SC-LDPC}-coded \ac{IDMA} systems with the windowed detection and decoding scheme as described in Sec.~\ref{sec:windowed_dec}. In the simulations, we use a decoding window length of $W_{d}=10$ sub-blocks, each consisting of 8000 symbols; $I_{max}=40$ iterations are carried out per window shift. As provided in Sec.~\ref{sec:de}, the code construction $C_1$ with $d_v=3$, $ d_c=6$, $d_{r}=4$, $W=3$ and the construction $C_2$ with $d_v=3$, $d_c=4$, $d_{r}=2$, $W=2$ are considered (for sum-rate one) and the total codeword length of the serially concatenated SC-LDPC and repetition code is fixed to $N_{\mathrm{CW}}=4\cdot 10^{5}$. The simulated BERs for both cases at $10^{-6}$ in AWGN channel are about $\unit[0.2]{dB}$ larger than the decoding thresholds obtained from density evolution. The gap to the ultimate GMAC capacity is about $\unit[1]{dB}$ and $\unit[1.7]{dB}$ for the (3,4) and (3,6) codes, respectively. Compared to the AWGN channel case, the performance loss due to Rayleigh fading is only $\unit[0.4]{dB}$ for both codes, due to the so-called multiuser diversity in fading channels \cite{DTseBookFund}. The performance of block LDPC code (code-word length is $N_{\mathrm{CW}}=4\cdot 10^{5}$, same as SC-LDPC codes) with EXIT chart based matching degree profiles is included for comparison. Besides the cumbersome matching procedure, the required irregularity in the parity-check matrix unfortunately degrades the error-floor performance of the code.

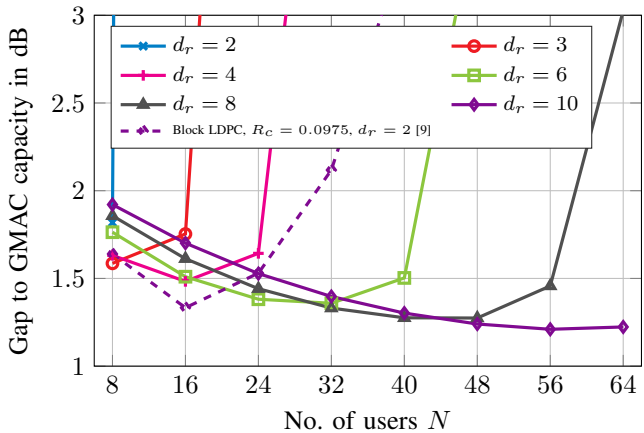
\begin{figure}
	\centering
	\begin{tikzpicture}

\begin{axis}[
width=\columnwidth,
height=6.25cm,
xmajorgrids,
yminorticks=true,
ymajorgrids,
yminorgrids,
legend pos=north west,        
legend style={legend cell align=left,align=left,draw=white!15!black, font=\footnotesize},
legend columns=2, 
legend style={
                    % the /tikz/ prefix is necessary here...
                    % otherwise, it might end-up with `/pgfplots/column 2`
                    % which is not what we want. compare pgfmanual.pdf
            /tikz/column 2/.style={
                column sep=5pt,
            },
},
xlabel={No. of users $N$},
ylabel={Gap to GMAC capacity in dB},
xtick={8,16,24,32,40,48,56,64},
%ymode=log,
%xmode=log,
mark size=2pt,
xmin=6,
xmax=66,
ymin=1,
ymax=3
]	
% \pgfplotstableread \datatable

\addplot[color=mittelblau, very thick,each nth point={1},mark=x] table [x index=0,y index =1] {data/Gap6.dat};
\addlegendentry{$d_{r}=2$}

\addplot[color= rot, very thick,each nth point={1},mark=o] table [x index=0,y index =1] {data/Gap5.dat};
\addlegendentry{$d_{r}=3$}

\addplot[color= pink, very thick,each nth point={1},mark=+] table [x index=0,y index =1] {data/Gap4.dat};
\addlegendentry{$d_{r}=4$}

\addplot[color= apfelgruen, very thick,each nth point={1},mark=square] table [x index=0,y index =1] {data/Gap3.dat};
\addlegendentry{$d_{r}=6$}

\addplot[color= dunkelgrau, very thick,each nth point={1},mark=triangle] table [x index=0,y index =1] {data/Gap2.dat};
\addlegendentry{$d_{r}=8$}

\addplot[color= lila, very thick,each nth point={1},mark=diamond] table [x index=0,y index =1] {data/Gap1.dat};
\addlegendentry{$d_{r}=10$}

\addplot[color= lila, dashed, very thick,each nth point={1},mark=diamond] table [x index=0, y expr=\thisrowno{1}+0.365] {data/LDPC_GA_rep_user_rep2.dat};
\addlegendentry{\tiny{Block LDPC, $R_{c}=0.0975$, $d_{r}=2$ \cite{wang2018idma}}}

%\addplot[color= orange, very thick, dashed,each nth point={1}] table [x index=0,y index =1] {tikz/Data/GA_rep_user_D.dat};
%\addlegendentry{designed}

\end{axis}

%\spy [black,width=2cm,height=2cm] on (spypoint1) in node [fill=none] at (magnifyglass1);
%\spy [black,width=2cm,height=1cm] on (spypoint2) in node [fill=none] at (magnifyglass2);s
\end{tikzpicture}
	\vspace*{-0.4cm}
	\caption{DE-based thresholds of the SC-LDPC $(d_v=3,d_c=6,L,W=3)$ code $C_1$  with varying number of users $N$ and repetition factor $d_{r}$.}
	\label{pgf:Gap}		
	%\vspace*{-0.7cm}
\end{figure}

As the number of users in a MAC system can vary, we propose to use the additional \ac{REP} to cope with the varying number of users \cite{wang2018idma}. Fig.~\ref{pgf:Gap} shows the \ac{GA} DE-based gap-to-capacity results for various repetition codes and number of users, while for all those curves the used \ac{SC-LDPC} $(d_v=3,d_c=6,L,W=3)$ code $C_1$ is fixed. It can be observed that increasing the repetition factor $d_{r}$ leads to a quite universal support over a wide range of number of users $N$. For instance with $d_{r}=10$, the gap-to capacity can be kept below $\unit[2]{dB}$ for the number of users in the range $N \in [8,64]$. For comparison, the results for a block LDPC code with optimized degree profile are also included, where the block LDPC code is of the rate $R_{c}=0.0975\approx 0.1$. With a further repetition code of $d_{r}=2$, the total code rate is the same as the (3,6) SC-LDPC code with $d_{r}=10$. It is obvious that (if fixed) the SC-LDPC code can support a wider range of users, although the gap-to-capacity can be larger for some number of users.  

\section{Conclusions and outlook}\label{sec:conclusions}
We have analyzed \ac{SC-LDPC} codes as coding scheme for an \ac{IDMA} multi-user system with a sliding windowed-based iterative detection and decoding receiver. We have examined the decoding thresholds through density evolution and shown that thresholds saturation occurs. 
The \ac{SC-LDPC} codes benefit from the anticipated \emph{universal} behavior of \ac{SC-LDPC} codes with respect to the channel front-end and, thus, do not require explicitly matched degree profiles for a specific number of users or changing channel characteristic. It also relaxes potential error-floor issues in finite-length code design, as regular node degrees instead of highly irregular degree profiles are sufficient.
Further, a windowed receiver implementation, consisting of both windowed detector and windowed decoder, keeps the overall decoding complexity within a feasible range but requires sub-block interleaving.
As a result, the proposed system operates below 1 dB away from the \ac{GMAC} capacity at a \ac{BER} of $10^{-6}$ for finite length code constructions.

\bibliographystyle{IEEEtran}
\bibliography{IEEEabrv,references}

\end{document}